\documentclass{emulateapj}

\usepackage{natbib}
\newcommand{\targst}{HD\,181327}
\newcommand{\refst}{$\alpha$ Ind}
\newcommand{\bQa}{\textit{Q$_a$}}

\begin{document}

\title{A Possible Icy Kuiper Belt around HD 181327}
\author{Christine H. Chen\altaffilmark{1}, Michael P. Fitzgerald\altaffilmark{2},
	\& Paul S. Smith\altaffilmark{3}}
\altaffiltext{1}{Spitzer Fellow; NOAO/STScI, Baltimore, MD 21218; cchen@stsci.edu}
\altaffiltext{2}{Michelson Fellow; Lawrence Livermore National Laboratory, L-413, Livermore, CA 94551}
\altaffiltext{3}{Steward Observatory, University of Arizona, Tucson, AZ 85721}

\begin{abstract}
We have obtained a Gemini South T-ReCS \bQa-band (18.3 $\mu$m) image and a \emph{Spitzer} MIPS SED-mode observation of \targst, an F5/F6V member of the $\sim$12 Myr old $\beta$ Pictoris moving group. We resolve the disk in thermal-emission for the first time and find that the northern arm of the disk is 1.4$\times$ brighter than the southern arm. In addition, we detect a broad peak in the combined \emph{Spitzer} IRS and MIPS spectra at 60 -75 $\mu$m that may be produced by emission from crystalline water ice. We model the IRS and MIPS data using a size distribution of amorphous olivine and water ice grains ($dn/da$ $\propto$ $a^{-2.25}$ with $a_{min}$ consistent with the minimum blow out size and $a_{max}$ = 20 $\mu$m) located at a distance of 86.3 AU from the central star, as observed in previously published scattered-light images. Since the photo-desorption lifetime for the icy particles is $\sim$1400 yr, significantly less than the estimated $\sim$12 Myr age of the system, we hypothesize that we have detected debris  that may be steadily replenished by collisions among icy Kuiper belt object-like parent bodies in a newly forming planetary system.
\end{abstract}


\section{Introduction}
Debris disks are dusty, gas-poor disks around main sequence stars. The estimated lifetimes for their constituent grains are typically shorter than the ages of their central stars by orders of magnitude, suggesting that the observed dust is debris, replenished from a reservoir such as by collisions among parent bodies or sublimation of comet-like bodies. To date, \emph{Infrared Astronomical Satellite (IRAS)} and \emph{Spitzer} studies have identified hundreds of debris disk candidates via thermal-emission from circumstellar dust (e.g. Hillenbrand et al. 2008; Rhee et al. 2007). Approximately 95\% of these candidates possess dust with cool color temperatures ($T_{gr}$ $<$ 100 K), similar to that expected for dust in the solar system's Kuiper Belt (Meyer et al. 2007).  Broad-band near-infrared photometry and low resolution spectroscopy suggest that KBOs possess ices on their surfaces (Schaller \& Brown 2007; Delsanti et al. 2006). The composition of dust in debris disks may be similar but has not yet been well-quantified (Chen et al. 2006).  

High resolution scattered-light and thermal-emission imaging and spectroscopy may reveal disk structure and provide insight into grain properties via (1) brightness peaks in scattered-light or thermal-emission images that indicate  local dust density enhancements and/or dust temperature asymmetries and (2) spectral features at mid- and far-infrared wavelengths. Only a handful of debris disks have been imaged both in thermal-emission and scattered-light thus far: $\beta$ Pictoris, HD 32297, HR 4796A, HD 107146, and Fomalhaut.  There may be tantalizing evidence to suggest that ices are present in these debris disks. Highly porous grains, similar to those used to model the infrared spectra of comets, have been used to model the $\beta$ Pictoris infrared spectrum and spectral energy distribution (Chen et al. 2007; Li \& Greenberg 1998). Excess thermal-emission in the inner region of the HD 32297 disk may indicate the presence of ice at temperatures near sublimation (Fitzgerald et al. 2007). However, moderate resolution, far-infrared spectra are needed to detect unambiguously emission from icy grains that is expected to be abundant in these systems. 

\targst\ is an F5/F6V star with a \emph{Hipparcos}-determined distance of 50.6 pc. It has been identified as a member of the $\beta$ Pictoris moving group, with an estimated age of $\sim$12 Myr (Zuckerman et al. 2001). Mannings \& Barlow (1998) first identified HD 181327 as a possible debris disk based on the detection of \emph{IRAS} 25 and 60$\mu$m excesses. \emph{Hubble Space Telescope (HST)} NICMOS and ACS coronagraphic imaging has resolved an inclined ring of dust in scattered-light at a distance of 86.3 AU with a width of 36 AU and an inclination, $i$ = 31.7$\arcdeg$ from face-on. The scattered light is $\sim$0.17\% of the incident stellar light and possesses a brightness asymmetry that is well fit using an asymmetric scattering parameter, $g$ = 0.30$\pm$0.03, at 1.1 $\mu$m, suggesting that the grains producing the scattered-light are small, with radii smaller than 1 $\mu$m (Schenider et al. 2006) if the grains are uniformly distributed in the disk. \emph{Spitzer} IRS observations measure a steeply rising 5-35 $\mu$m spectrum that is well characterized by an 80 K single-temperature black body and a fractional infrared luminosity, $L_{IR}/L_{*}$ = 3.1$\times$10$^{-3}$ (Chen et al. 2006), commensurate with that measured toward $\beta$ Pictoris. We have carried out a high resolution multi-wavelength imaging and spectroscopy study to infer the grain properties around \targst.

\section{Observations}

\subsection{T-ReCS Observations}
We used the T-ReCS imager on Gemini-S to observe \targst\ on the night of 2006 July 21 (GS-2006A-Q-43).  These observations were interleaved with exposures of \refst, a photometric and point-spread function (PSF) reference.
Both targets were imaged in the \bQa\ band ($\lambda_0=18.3$\,\micron) with an ABAB chop-nod sequence (15\arcsec-throw at P.A. = 30\degr).
The total guided, on-source exposure times were 4865\,s and 456\,s for \targst\ and \refst, respectively.

The data were reduced using custom Python routines.  In order to suppress sky and instrumental emission, we used standard addition and subtraction tehcniques for combining the exposures in each AB nod pair.
For \targst, two of the resulting double-difference images with abnormally high variances (more than twice the median value) were discarded, reducing the on-source exposure time by 43\,s.  Each of the remaining nod-pair images was then weighted by its variance, and images with common pointings were summed.
These resulting images were then processed to remove electronic offsets.  After masking a 3\arcsec-radius region around the target, the median values in each row and (20-column) amplifier channel were used to iteratively solve for and remove offsets.
Each image was then registered at the one-pixel level with Gaussian centroids,
and the results were shifted and summed to a single image.
This image was similarly processed to remove offsets in each row and column (as opposed to each row and 20-column channel in the earlier step).

\begin{figure}
\plotone{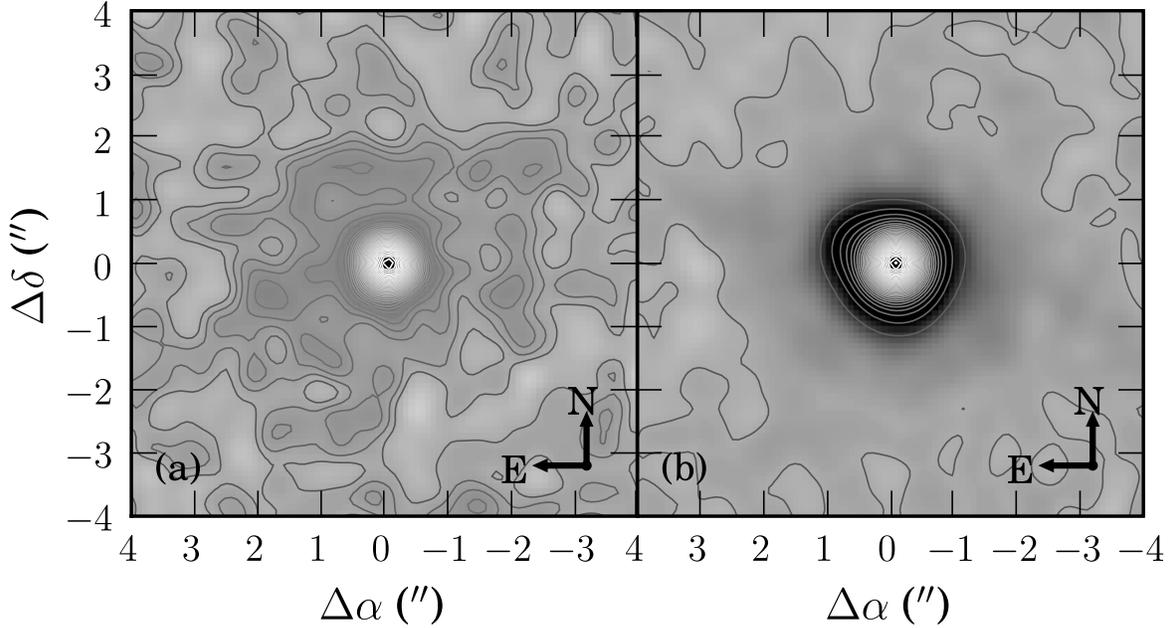}
\caption{\textit{(a)} \bQa-band observation of \targst.  Contours are spaced according to the smoothed 1-$\sigma$ background noise level.  \textit{(b)} PSF and photometric reference star \refst, with contours scaled to match the brightness levels in panel \textit{(a)}.  Both images' contours have been smoothed by a FWHM=0\farcs5 Gaussian.}\label{fig:images}
\end{figure}

Detailed examination of the resulting image of \targst\ revealed additional electronic pattern noise, seen as diagonal banding.
The power spectrum of the image contained peaks at $(n\nu_x,\nu_y)$, where $\nu_x^{-1}=20$\,pix (corresponding to the amplifier channel width), $\nu_y^{-1}=55$\,pix, and $n=1,2,\cdots$.
After masking a 3\arcsec-radius region around the star, Hermite-symmetric Fourier modes at spatial frequencies within 1.5\,pix$^{-1}$ of the $n=1$--3 peaks were fit to the image of \targst\ and subtracted.
Additionally, all power at frequencies above the diffraction limit $\nu_\mathrm{tel}=D/\lambda_\mathrm{min}$ (taking $\lambda_\mathrm{min}=17.57$\,\micron\ for \bQa) was removed from the images of both \targst\ and \refst.  The results of this process are shown in Figure~\ref{fig:images}.
While the subtraction of Fourier modes at the power spectrum peaks reduced the 1-$\sigma$ noise level by 1.4\% (3.3\% in smoothed images), the primary purpose of this step was to remove spatial correlations in the noise which might masquerade as faint circumstellar emission.

We used aperture photometry of \refst\ for photometric calibration.
We computed the flux zero point using an aperture of 3\farcs6 radius, and a \bQa\ flux value of 6.032\,Jy.
An examination of the enclosed flux as a function of airmass showed no clear trend, suggesting time-variable telluric extinction.
The mean airmasses of the \targst\ and \refst\ exposures were 1.18 and 1.22, respectively.
We elected not to apply a correction for the expected difference in mean extinction between the two stars.  Rather, we note that time-varying extinction is likely to dominate the systematic photometric errors.  We note that aperture photometry of \refst\ in nod-pair difference images had a variation of 11\%, which we take as an estimate of the systematic uncertainty in the photometric zero point.

\subsection{MIPS SED-mode Observations}
We obtained follow-up \emph{Spitzer} (Werner et al. 2004) MIPS (Rieke et al. 2004) SED-mode observations (55.0--90.0 $\mu$m; $\lambda/\Delta \lambda \sim$ 20) of \targst\ (AOR key: 16171008) on 15 April 2006, using the 19.6$\arcsec$$\times$157$\arcsec$ slit/grating and 10 cycles of 10 s integrations. Each observing cycle consists of six pairs of 10 s exposures with the slit position alternated between the object and blank sky 1$\arcmin$ away. To remove the background, each sky exposure is subtracted from the immediately preceding frame containing the object. Exposures of an internal calibration source are interspersed within the cycle to track the varying response of the 70 micron detector array. The raw MIPS data were corrected for distortion, registered, mosaicked, and flat fielded using the MIPS instrument team's data analysis tool (DAT; Gordon et al. 2005). The spectrum was extracted using a 5-pixel extraction aperture and calibrated using the procedure and calibration files described in Lu et al. (2008). The processing steps include applying the dispersion solution, an aperture correction (to account for slit losses) and a flux calibration based on $\sim$20 IR standard stars (Gordon et al. 2007).

We use these newly obtained MIPS SED-mode data in conjunction with previously published \textit{Spitzer}/IRS observations to model the SED of \targst. The observations and data reduction for the IRS data are described in Chen et al. (2006). We estimate the \targst\ stellar photosphere using a Kurucz stellar atmosphere model with $T_\mathrm{eff}$ = 6500 K, $\log g$ = 4.0, and solar metallicity, normalized to the first 10 data points ($\lambda$ $<$ 6 $\mu$m) in the IRS SL spectrum. We assume a 5\% uncertainty in our stellar photosphere model.

\section{Thermal Imaging Results}

\begin{figure}
\plotone{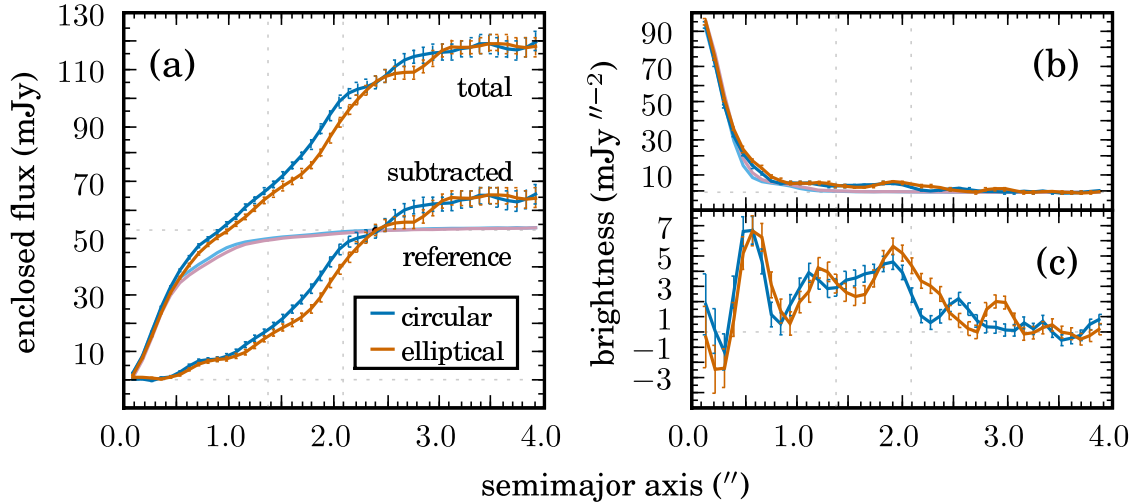}
\caption{\textit{(a)} Enclosed flux curves for circular and elliptical ($e=0.525$, P.A.$=107$\degr) apertures.  The curve labeled ``total'' corresponds to the image in Fig.~\ref{fig:images}a, while the ``reference'' curve is derived from the PSF star \refst, scaled to the estimated photospheric flux ($F_{\nu,*}=53$\,mJy) of \targst.  The ``subtracted'' curve is the difference between the two, showing a detected \bQa-band excess of 61.5$\pm$2.7\,mJy inside a 3\arcsec\ circular aperture. \textit{(b)} Azimuthally averaged brightness profiles derived from the enclosed flux curves for \targst\ and \refst\ (scaled to $F_{\nu,*}$).  \textit{(c)}  Same as \textit{(b)} from the PSF-subtracted curve in panel \textit{(a)}.  All panels show vertical lines marking $e$-folding distances which bracket the peak of the \textit{HST}/NICMOS scattered-light profile~\citep{sch06}.  Error bars are derived from background noise levels, and do not include systematic errors in the flux zero point, and subtracted profiles do not include PSF uncertainty.}\label{fig:profiles}
\end{figure}

We plot our \bQa-band images of \targst\ and PSF reference star \refst\  in Figure~\ref{fig:images},
and their enclosed flux curves in Figure~\ref{fig:profiles}a, with 1-$\sigma$ uncertainties derived from the background noise.
We subtract the PSF star image from our HD 181327 image, assuming that the stellar point source has a flux of 53 mJy (estimated from our stellar photosphere model), revealing thermal-emission from the disk within apertures $\lesssim$3\arcsec (Figure 3).  The lack of excess interior to 0\farcs4 suggests that our photospheric flux estimate is accurate.
We measure an excess \bQa-band flux of 61.5$\pm$2.7\,mJy within a 3\arcsec\ circular aperture,  consistent with our analysis of the \textit{Spizter}/IRS spectrum.

\begin{figure}
\plotone{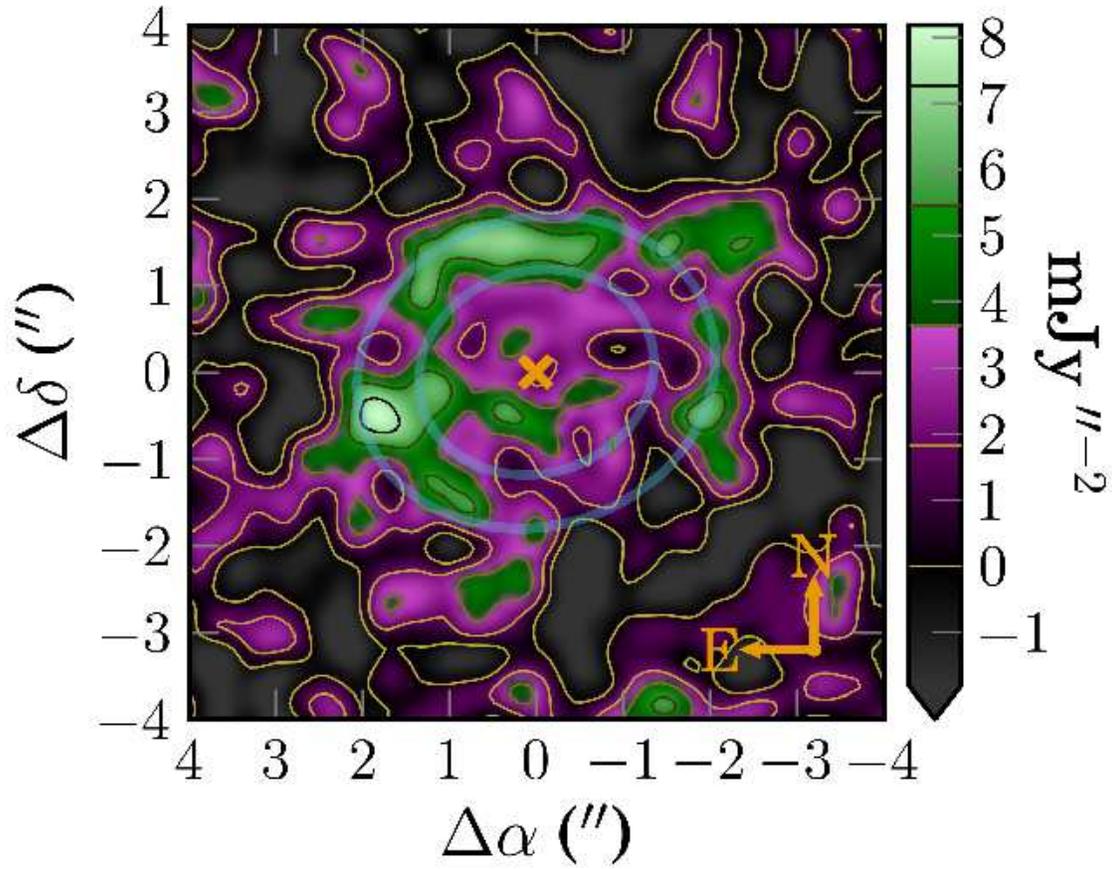}
\caption{Residual image of \targst\ after subtraction of the stellar PSF.  The image here has been smoothed by a FWHM=0\farcs5 Gaussian.  Contours are spaced according to the 1-$\sigma$ smoothed background noise level.  The magenta/green transition occurs at the 2-$\sigma$ level.  Ellipses show the $e$-folding distances which bracket the peak of the \textit{HST}/NICMOS scattered-light profile~\citep{sch06}.}\label{fig:psfsub}
\end{figure}

\textit{HST} ACS and NICMOS scattered-light imaging resolve an inclined ring of dust around \targst\ with (1) a surface brightness that falls off rapidly interior to the 86.3\,AU ring and (2) an azimuthal brightness asymmetry that is well-modeled using a Henyey-Greenstein phase function. These observations suggest that the inner region of the disk possesses little dust and assume that the distribution of dust in the ring is azimuthally symmetric.
In Fig.~\ref{fig:psfsub}, we indicate the extent of the scattered-light ring with ellipses corresponding to the inner and outer $e$-folding radii of the deprojected scattered-light profile~\citep[][\S5.2]{sch06}.  

The northern-arm of the disk appears brighter than the southern arm, contrary to the equal brightness expected from an azimuthally-symmetric dust ring. We measure the flux in annular wedges, centered on the minor axis, with outer boundaries defined by the outer $e$-folding radius measured in scattered light. We measure a fluxes of 18 $\pm$ 1 mJy and 13 $\pm$ 1 mJy in a 120$\arcdeg$ wedges in the northern and southern arms, respectively, corresponding to a flux ratio of 1.4 $\pm$ 0.1 between the northern and southern arms. Therefore, the observed large scale azimuthal variation in scattered light may not be purely generated by the dust scattering phase function as previously assumed. Instead, the dust density and/or temperature may be higher in the northern arm than in the southern arm. Faint diffuse emission is observed in scattered-light with ACS on the northern side of the disk just outside the main dusty ring (Schneider et al. 2006). This diffuse emission may trace small grains that must be replenished because they are being radiatively driven from the system. One possible scenario is that the northern \bQa-band enhancement is the site of recent collisional activity; however, small grain structures are expected to disappear on 1000 year timescales either because small grains are removed by radiation pressure or they are smeared out azimuthally by orbital motion. 

If the dust grain density is higher in the northern arm, then previous models of the scattering phase function may need to be revised. Schneider et al. (2006) fit the 1.1 $\mu$m NICMOS scattered-light image of the \targst\ disk using a Henyey-Greenstein-like phase function with an asymmetric scattering parameter, $g$ = 0.30 $\pm$ 0.03, assuming that the grains are uniformly distributed in azimuth. This model implies that the scattered light from the northern 120$\arcdeg$ wedge is 2.1 $\pm$ 0.1 times brighter than the scattered light from the southern one. If, however, the grain density in the northern 120$\arcdeg$ wedge is 1.4 $\pm$ 0.1 times larger then that in the southern one,  then grains on the northern side forward scatter (2.1/1.4=) 1.5$\times$  the amount of light compared to grains on the southern side, corresponding to a revised, $g$ = 0.16$\pm$0.04. Since the physical origin of the thermal brightness enhancement on the northern side is currently unknown, 0.12 $<$ $g$ $<$ 0.33.

Our PSF-subtracted $Q_{a}$-band image shows possible emission interior to the 86.3\,AU ring; however, this structure may have been generated by changes in the PSF structure during the 5.5 hours over which our observations were made. Warm dust components have been observed interior to the dusty rings around HR 4796A and Fomalhaut. Koerner et al. (1998) and Stapelfeldt et al. (2004) have argued that these grain populations may be generated by collisions among asteroid-like parent bodies. High resolution thermal-emission imaging at shorter wavelengths is needed to confirm the presence of a possible warm dust component around \targst\ at higher spatial resolution.

\section{A Two Grain Model}
\subsection{Thermal-Emission}
\begin{figure}
\plotone{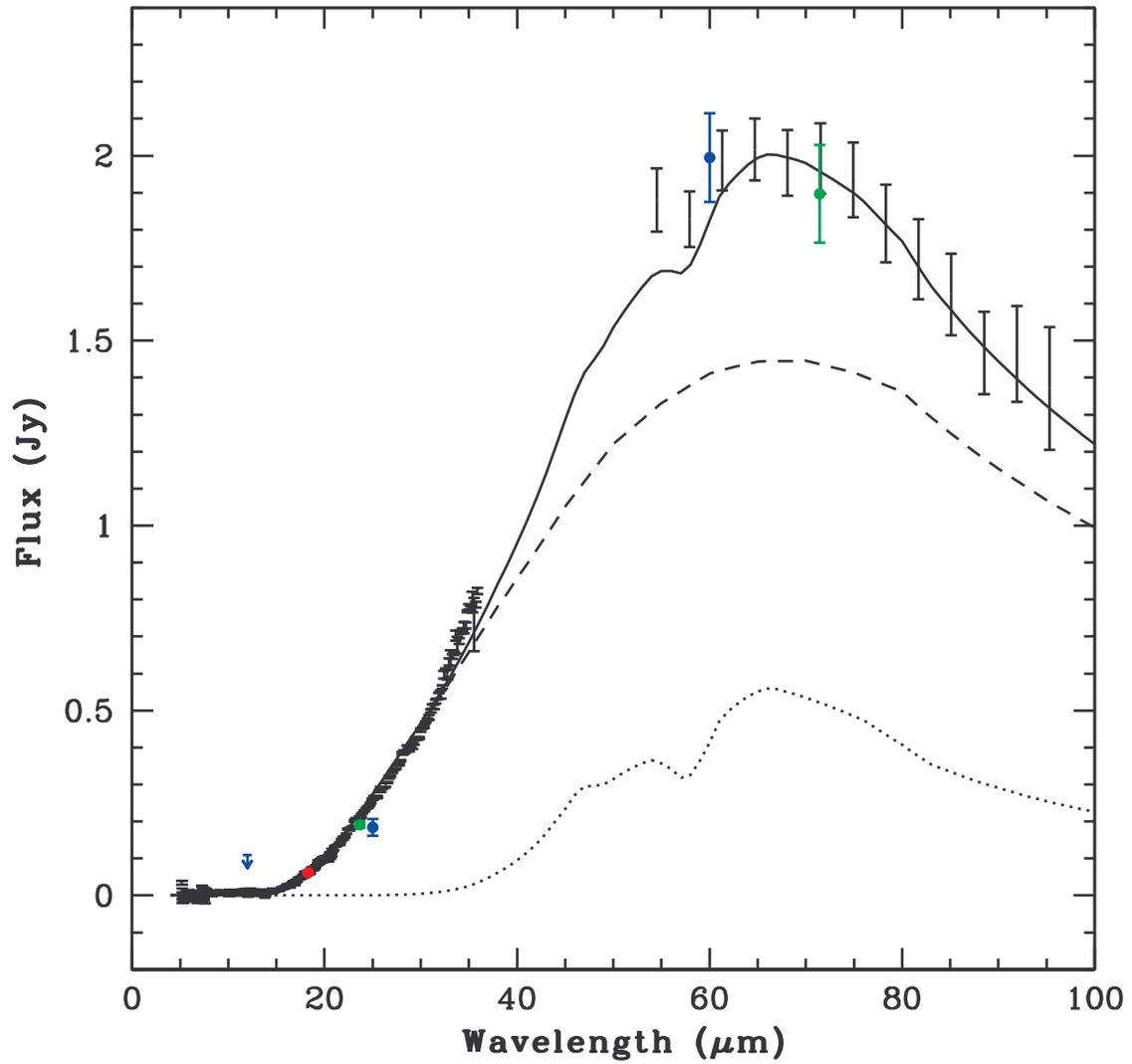}
\caption{Photosphere-subtracted IRS and SED-mode spectra for HD181327, shown with black error bars. Overlaid are T-ReCS (this work), MIPS (Smith et al. 2006), and \emph{IRAS} photometry (not color-corrected), shown with red, green, and blue error bars. Model emission from amorphous olivine and crystalline water ice are shown with dashed and dotted lines, respectively. The final model for the system is shown with a solid black line.}\label{fig:spectra}
\end{figure}

Single-temperature black bodies grains and power-law size distributions of  olivine grains are inadequate for reproducing the observed HD 181327 grain properties. (1) Black body grains in radiative equilibrium, located at a distance of 86 AU around a $L_{*}$ = 3.1 $L_{\sun}$, $T_{eff}$ = 6450 K star (Schenider et al. 2006), are expected to possess a grain temperature, $T_{gr}$ = 40 K, significantly cooler than that inferred from the infrared SED (80 K). (2) Single temperature black body fits to the IRS and MIPS SED-mode data are unable to reproduce the overall shape of the spectrum. The photosphere-subtracted SED for \targst\ (Fig.~\ref{fig:spectra}) possesses a sharp rise in the flux at wavelengths $>$20 $\mu$m that peaks at 60-75 $\mu$m. Adequate fits to IRS continuum produce only three quarters of the observed 60-75 $\mu$m flux. (3) Power-law size-distributions of olivine grains (with $a_{min}$ determined by the minimum blow-out size and $a_{max}$ = 20 $\mu$m) that successfully reproduce the gross features of the HD 181327 SED scatter 0.2 mJy arcsec$^{-2}$ of the incident stellar light at 1.1 $\mu$m and 90$\arcdeg$ scattering, approximately 25\% of that observed. 

While the MIPS SED-mode data have very low resolution (R $\sim$ 15-25) and do not cover the 40 - 55 $\mu$m range, the 60-75 $\mu$m peak indicates emission from a relatively broad solid state feature. Since water ice is expected to be abundant and crystalline water ice possesses a well-studied emission plateau at 40 - 70 $\mu$m with features at 43 and 62 $\mu$m (Omont et al. 1990), we model the far-infrared SED using amorphous olivine grains to create the underlying continuum and water ice (polluted with 0.1\% amorphous carbon by volume) to reproduce the observed emission features. We calculate absorption coefficients using laboratory measurements of the optical constants for amorphous olivine (MgFeSiO$_{4}$; Dorschner et al. 1995), water ice near freezing (Warren 1984), and amorphous carbon (Preibisch et al. 1993) using Bruggeman effective medium and Mie theory (Bohren \& Huffman 1983). 

We model the observed SED by summing over the thermal-emission produced by each species, requiring that both the amorphous olivine and water-ice grains are located at 86.3 AU, consistent with the 1.1 $\mu$m scattered-light image
\begin{equation}
F_{\nu} = \sum_{s} \int_{a_{min}}^{a_{max}}  \frac{\pi a^2}{d^2} \frac{dn}{da} Q_{\nu,abs,s} B_{\nu}[T_{s}(a)] da,
\end{equation}
where $a$ is the grain radius, $dn/da \propto a^{-p}$ is the grain size distribution with minimum grain radius, $a_{min}$, and maximum grain radius, $a_{max}$; $Q_{\nu,abs,s}$ and $T_{s}$ are the absorption efficiency and temperature of grain species, $s$, and $d$ is the distance from the observer to the star. We estimate the minimum grain size assuming that grains with $\beta$ = $F_{rad}/F_{grav}$ $<$ 0.5 will be gravitationally bound:
\begin{equation}
a_{min} = \frac{6 L_{*} \langle Q_{pr,s} \rangle}{16 \pi G M_{*} \rho_{s}}
\end{equation}
(Burns et al. 1979; Artymowicz 1988), where $L_{*}$ (= 3.1 $L_{\sun}$) and $M_{*}$ (= 1.4 $M_{\sun}$) are the luminosity and mass of HD 181327, $\langle Q_{pr,s}(a) \rangle$ = $(\int Q_{pr}(a,\lambda) F_{\lambda} d\lambda)$ $(\int F_{\lambda} d\lambda)^{-1}$ is the radiation pressure efficiency averaged over the stellar spectrum, and $\rho_{s}$ is the species density. For HD 181327, we estimate minimum grains sizes, $a_{min}$ = 1.0 and 1.5 $\mu$m for amorphous olivine ($\rho_{s}$ = 3.71 g cm$^{-3}$) and water ice ($\rho_{s}$ = 0.92 g cm$^{-3}$), respectively. We set $a_{max}$ = 20 $\mu$m. Since $2 \pi a_{max}$ $>$ 100 $\mu$m, the presence of larger grains should not dramatically impact our SED model. We let $p$ be a free parameter and require that $p$ be the same for both the olivine and ice distributions.

We reproduce the observed SED using 6.4$\times$10$^{25}$ g amorphous olivine grains with $dn/da$ $\propto$ $a^{-2.25}$ (with estimated $T_{gr}$ = 48 - 93 K) and 2.0$\times$10$^{25}$ g of  water ice grains also with $dn/da$ $\propto$ $a^{-2.25}$ (with estimated $T_{gr}$ = 39 - 50 K; see Fig.~\ref{fig:spectra}). Our model SED possesses a reduced $\chi^{2}$ value of 22; it is slightly deficient in flux at wavelengths 30 $\mu$m $<$ $\lambda$ $<$ 55 $\mu$m. Our T-ReCS image raises the possibility that there may be additional dust grains interior to the ring that we have not included in our model. If such warm dust grains exist, then their emission may account for the discrepancy between our observed and synthetic SEDs. Numerical simulations of collisions among dust grains around early-type stars suggest that radiation pressure modifies the size distribution from theoretical collisional equilibrium by removing grains with sizes just above the blow-out size (Th\'{e}bault \& Augereau  2007; Krivov et al. 2006). The shallow grain size distribution inferred from our modeling may indicate that small grains adjacent to the blow-out size have been removed by radiation pressure. Indices of refraction have been measured for water ice created at temperatures lower than freezing in which the ice is more amorphous (Hudgins et al. 1993), and the 42 and 63 $\mu$m emission features for 100 K amorphous water ice  are significantly less pronounced than for water ice near freezing. The overall better fit provided by crystalline water ice may indicate that the ice has been heated in the past and has since cooled to the present temperature. 

\subsection{Scattered-Light}
We estimate the scattered-light surface brightness, generated by our grain population, with scattering efficiency, Q$_{\nu,sca,s}$, radius, $a$, and size column density distribution, $dN_{gr,s}/da$,
\begin{equation}
SB = \sum_{s} \frac{L_{*,\nu}}{4 \pi D^2} 
	\int_{a_{min}}^{a_{max}} f_{s}(\theta) Q_{\nu,sca,s} \pi a^2 \frac{dN_{gr,s}}{da} da 
\end{equation}
(Augereau \& Beust 2006), where L$_{*,\nu}$ is the specific luminosity of the star, $D$ is the distance to the grains, and $f_{s}(\theta)$ is the scattering phase function of species, $s$. The specific luminosity of the star can be inferred from its observed flux and distance, $L_{*,\nu} = 4 \pi d^{2} F_{*,\nu}$, where $d = 50.6$\,pc is the stellar distance and $F_{*,\nu}(1.1\,\micron)=5.9$\,Jy (Schneider et al. 2006) is the inferred stellar flux in the NICMOS filter. The scattering phase function can be approximated by the Henyey-Greenstein phase function,
\begin{equation}
f_{s}(\theta) = \frac{1-g_{s}^2}{4 \pi} \frac{1}{(1 + g_{s}^2 - 2g_{s}\cos{\theta})^{3/2}}
\end{equation}
(Henyey \& Greenstein 1941), where the scattering angle, $\theta$, is the angle of deviation from forward scattering. If the dust is azimuthally symmetric, then we estimate a 1.1 $\mu$m scattered-light surface brightness, SB = 0.81 mJy arcsec$^{-2}$ at $D_{o}$ = 86.3 AU and $\theta$ = 90$\arcdeg$ (the disk ansa), consistent with observations by Schneider et al. (private communication).

In addition to the scattered-light surface brightness, the ACS and NICMOS observations make two additional measurements of the grain population around HD 181327. Schneider et al. (2006) report a scattered-light disk color [F606W]-[F110W] = 0.5 $\pm$ 0.3 mag at $D_{o} = 86.3 AU$. The dust grains in our model possess a color [F606W]-[F110W] = -0.07 mag, 1.9$\sigma$ lower. Schneider et al. (2006) also report an asymmetric scattering factor $g$ = 0.30$\pm$0.03 which we revise to 0.12 $<$ $g_{HG}$ $<$ 0.33 in our discussion of the resolved thermal-emission (\S3). The dust grains in our model possess  $\langle g \rangle$ = 0.87, substantially higher than inferred from the scattered light observations, because water ice forward-scatters light efficiently independent of grain size. The large discrepancy between the observed and inferred asymmetric scattering parameters may be the result of the unrealistic assumption of spherical grains. Models including more realistic grain shapes may decrease the inferred value of $g$ while reproducing the overall thermal emission and scattered light from the disk. More detailed modeling of the dust grains in this system is needed to reproduce all of the observations.

\section{Discussion}

Recent theoretical work suggests that $\mu$m-sized water ice grains around A- and F-type main sequence stars are more effectively destroyed via photo-desorption by stellar ultra-violet photons than by sublimation (Grigorieva et al. 2007). We estimate the water ice grain erosion rate due to photo-desorption
\begin{equation}
\dot{a} = - \frac{\eta m_{(H2O)} Y N_{abs}} {4 \rho_{i}}
\end{equation}
where $\eta$ is the fraction of the surface covered by ice, $m_{(H2O)}$ is the mass of a water molecule (= 3$\times$10$^{-23}$ g), $\rho_{i}$ is the density of water ice (=0.92 g cm$^{-3}$), $Y$ is the desorption efficiency at ultraviolet wavelengths (=10$^{-3}$), and $N_{abs}$ is the flux of absorbed photons at the location of the water ice grains,
\begin{equation}
N_{abs} = \int_{\lambda_{min}}^{\lambda_{max}}
\frac{\lambda L_{\lambda}}{4 \pi D^2 hc} Q_{abs}(\lambda) d\lambda,
\end{equation}
where $L_{\lambda}$ is the specific stellar luminosity, $\lambda_{min}$ (= 0.091 $\mu$m), and $\lambda_{max}$ (= 0.24 $\mu$m) (Grigorieva et al. 2007). We estimate that the flux of absorbed ultra-violet photons that desorb water molecules from ice grains, $N_{abs}$ = 4.3$\times$10$^{11}$ photon s$^{-1}$ cm$^{-2}$, at $D_{o}$ = 86.3 AU, assuming \targst\ is 10$\times$ brighter at ultra-violet wavelengths than inferred from the stellar photosphere model, consistent with TD1 observations (Thompson et al. 1978). We calculate $\dot{a}$ = 1.1 $\times$10$^{-3}$ $\mu$m yr$^{-1}$ if the surface of the grains are 99.9\% water ice and 0.1\% amorphous carbon ($\eta$ = 0.999), corresponding to a photo-desorption lifetime of 1400 yr for grains with radii, $a$ = 1.5 $\mu$m, similar to the estimated collision timescale for grains in this system ($\sim$1300 yr; Chen et al. 2006). Since the photo-desorption timescale is significantly shorter than the stellar age ($\sim$12 Myr), we hypothesize that icy grains are replenished through collisions among icy parent bodies, similar to Kuiper Belt objects in our solar system.

The cold grain temperatures ($T_{gr}$ $<$ 100 K; Meyer et al. 2007) in debris disks has led to speculation that these systems are icy Kuiper Belt analogs; however, the difficulty in detecting mid-infrared spectral features from these systems (Chen et al. 2006; Jura et al. 2004) makes confirmation of this supposition challenging. To date, the most evolved young, circumstellar disk with observed water ice absorption/emission features is the isolated, Herbig F7IIIe star HD 142527. \emph{ISO} SWS and LWS spectra of this source detect both the 43 and 62 $\mu$m emission features associated with the presence of cold, predominantly crystalline, water ice (Malfait et al. 1999). Debris disks around main sequence stars are fainter than their younger Herbig Fe counterparts, making detection of water ice emission even more challenging. The possible evidence indicating the presence of water ice emission around HD 181327 is the first for a debris disk. Approximately 25\% of the cold dust around \targst\ may be water ice, significantly less than the 60\% observed toward HD 142527. The relative deficiency of water  ice around \targst\ may indicate that it has been incorporated into larger bodies (that are less efficient emitters) and/or removed via photo-desorption and radiation pressure. 

\section{Conclusions}
We have obtained a T-ReCS \bQa-band (18.3 $\mu$m) image and a  55 - 90 $\mu$m MIPS SED-mode spectrum of \targst, an F5/F6V star in the $\beta$ Pic moving group with a resolved scattered-light disk. Based on our analysis of the multi-wavelength imaging and spectroscopy, we conclude the following:

1. The northern arm of the \targst\ disk is 1.4 times brighter than the southern arm in thermal-emission, suggesting that either the density and/or the temperature of the grains in this region is higher than in the southern arm of the disk.

2. The overall properties of the unresolved spectral energy distribution (at 5.5 - 90 $\mu$m), the resolved thermal-emission at 18.3 $\mu$m, and the resolved scattered-light at 1.1 $\mu$m can be reproduced by a population of 1 - 20 $\mu$m amorphous olivine and crystalline water ice grains located at a distance of 86.3 AU from the central star.

3. Since the estimated photo-desorption lifetime of 1.5 $\mu$m water ice grains is 1400 yr, significantly shorter than the age of \targst, the grains must be replenished from a reservoir such as collisions among parent bodies, perhaps Kuiper Belt objects. 

\acknowledgements
We would like to thank J. Kessler-Silacci and A. Noriega-Crespo for their assistance with the preliminary SED-mode data reduction and S. Sandford for providing an electronic copy of the laboratory measured optical constants of ices published in Hudgins et al. (1993). We would also like to thank J. Debes, M. Jura, C. McCabe, J. Najita, D. Watson, A. Weinberger, K. Willacy, and our anonymous referee for their helpful comments and suggestions. Support for this work at NOAO/STScI was provided by NASA through the \emph{Spitzer Space Telescope} Fellowship Program, through a contract issued by the Jet Propulsion Laboratory, California Institute of Technology under a contract with NASA. M.P.F. acknowledges support from the Michelson Fellowship Program, under contract with JPL, funded by NASA. Work at LLNL was performed under the auspices of DOE under contract DE-AC52-07NA27344. Support for this work at Steward Observatory was provided by JPL/\emph{Spitzer} contract 1256424.


%


\end{document}